\documentclass[11pt]{article}
\usepackage{graphicx}

\newcommand{\B}{B^1}

\newcommand{\qR}{q^1_R}

\def\singleandabitspaced{\baselineskip=\normalbaselineskip\multiply
    \baselineskip by 110\divide\baselineskip by 100}
\def\singlespaced{\baselineskip=\normalbaselineskip}

\newcommand{\centeron}[2]{{\setbox0=\hbox{#1}\setbox1=\hbox{#2}\ifdim
                             \wd1>\wd0\kern.5\wd1\kern-.5\wd0\fi \copy0
                             \kern-.5\wd0\kern-.5\wd1\copy1\ifdim\wd0>\wd1
                             \kern.5\wd0\kern-.5\wd1\fi}}
\newcommand{\ltap}{\>\centeron{\raise.35ex\hbox{$<$}}
                     {\lower.65ex\hbox{$\sim$}}\>}
\newcommand{\gtap}{\>\centeron{\raise.35ex\hbox{$>$}}
                     {\lower.65ex\hbox{$\sim$}}\>}

\newcommand{\lsim}{\mathrel{\ltap}}

\begin{document}

\singlespaced

\begin{titlepage}

\begin{center}
\vspace*{0.8in}
\mbox{\Large \textbf{Searching For Dark Matter 
with Neutrino Telescopes}} \\
\vspace*{1.6cm}
{\large Dan Hooper$^1$ and Joseph Silk$^{1,2}$} \\
\vspace*{0.5cm}
$^1$ {\it Denys Wilkinson Laboratory, Astrophysics Department, OX1 3RH Oxford, England, UK} \\
$^2$ {\it Institut d Astrophysique de Paris, Paris, France} \\
\vspace*{0.6cm}
{\tt hooper@astro.ox.ac.uk, silk@astro.ox.ac.uk} \\
\vspace*{1.5cm}
\end{center}

\begin{abstract} 
\singleandabitspaced

\end{abstract}

One of the  most interesting mysteries of astrophysics is the puzzle of dark matter. Although numerous techniques have been explored and developed to detect this elusive substance, its nature remains unknown. One such method uses large high-energy neutrino telescopes to look for the annihilation products of dark matter annihilations. In this summary article, we briefly review this technique. We describe the calculations used to find the rate of capture of WIMPs in the Sun or Earth and the spectrum of neutrinos produced in the resulting dark matter annihilations. We will discuss these calculations within the context of supersymmetry and models with universal extra dimensions, the lightest supersymmetric particle and lightest Kaluza-Klein particle providing the WIMP candidate in these cases, respectively. We will also discuss the status of some of the experiments relevant to these searches: AMANDA, IceCube and ANTARES.

\end{titlepage}

\newpage
\setcounter{page}{2}
\singleandabitspaced

\section{Introduction}

There exists an enormous body of evidence in the favor of cold
dark matter. This evidence includes observations of galactic
clusters and large scale structure \cite{structure}, supernovae
\cite{supernovae} and the cosmic microwave background (CMB) anisotropies
\cite{cmb,wmap}. Recently, WMAP has provided the most detailed
information on the CMB to date, quoting a total matter density of
$\Omega_{\rm{m}} h^2 = 0.135^{+0.008}_{-0.009}$
\cite{wmap}. Furthermore, data from WMAP and other prior experiments
indicate a considerably smaller quantity of baryonic matter
\cite{wmap,nonbaryonic}. At the 2-$\sigma$ confidence level, the
density of non-baryonic, and cold dark matter is now known to be
$\Omega_{\rm{CDM}} h^2 = 0.113^{+0.016}_{-0.018}$ \cite{wmap}. 

Many methods have been proposed to search for evidence of particle
dark matter. In addition to accelerator searches, direct and indirect dark matter detection experiments
have been performed. Direct dark matter searches attempt to measure
the recoil of dark matter particles scattering elastically off of the
detector material \cite{direct}. Indirect dark matter
searches have been proposed \cite{ss}
 to observe the products of dark matter annihilation
including gamma-rays
\cite{indirectgamma,subgamma}, positrons \cite{positrons}, anti-protons \cite{antiprotons} and neutrinos \cite{indirectneutrino}.
 
Each of these methods has its advantages and disadvantages. Direct detection is most suited for WIMPs with relatively large elastic scattering cross sections with nucleons. Indirect detection using gamma-rays is often studied in the context of observing annihiltions in the galactic center region, which depends dramatically on the dark matter profile used. Alternatively, dark matter substructure could be studied with this method. This would still depend on poorly known dark matter distributions, however \cite{subgamma}. Measurements of the local anti-proton or positron cosmic-ray spectrum depends critically on the location of, and amount of, dark matter substructure within tens of kiloparsecs and a few kiloparsecs, respectively. Also, positrons and anti-protons do not point towards their sources making an unambiguous separation from backgrounds very difficult.

WIMPs which scatter elastically in the Sun or Earth may become gravitationally bound in these gravitational wells. Over the age of the solar system, they may accumulate in substantial numbers in these objects, greatly enhancing their annihilation rate. Although gamma-rays, positrons and anti-protons produced in these annihilations do not escape the Sun or Earth, neutrinos often can. Also, unlike other indirect dark matter searches this method does not depend strongly on our galaxy's dark matter halo profile or on the distribution of dark matter substructure. In this article, we review this method of indirect dark matter detection, and discuss the prospects for such observations in existing and planned experiments.

\section{Particle Dark Matter Candidates}

Numerous particles have been discussed in the literature as dark matter candidates. It would be impossible, and undesirable, for us to attempt to review them all here. Instead, we will focus on two examples which arise in popular particle physics scenarios: supersymmetry and models with universal extra dimensions. 

\subsection{Supersymmetry}

In models with supersymmetry, each fermion has a bosonic partner (and vice versa) which exactly cancels all quadratic divergences to the Higgs mass, thus naturally solving one of the major problems associated with the standard model. Furthermore, the lightest supersymmetric particle (LSP), is stable in most viable models due to the conservation of R-parity \cite{rparity}. In many supersymmetric models, the LSP is the lightest neutralino, a mixture of the superpartners of the
photon, Z and neutral higgs bosons, and is electrically
neutral, colorless and a viable dark matter candidate \cite{susylsp}.   

If such a particle were in equilibrium with photons in the early universe,
as the temperature decreased, a freeze-out would occur, leaving a thermal
relic density. The temperature at which this occurs, and
the density which remains, depends on the annihilation cross section
and mass of the lightest neutralino. It is natural for supersymmetry to provide
a dark matter candidate with a present abundance similar to those
favored by the WMAP experiment.  

The details of any SUSY model are subject to the way supersymmetry is broken (i.e. how masses are given to the superpartners). Most often, the LSP is predominantly-gaugino, although it is possible to have a substantial higgsino fraction in some cases.

\subsection{Universal Extra Dimensions}

Models with extra dimensions appearing at or near the TeV scale have become very popular in recent years. One class of these models are those in which all of the fields of the standard model propagate in ``universal'' extra dimensions \cite{antoniadis,ACD}. In these models, Kaluza-Klein (KK) excitations appear as particles with masses of the scale of the extra dimension. Due to conservation of momentum in the higher dimensions, a symmetry called KK-parity can arise which can, in some cases, make the Lightest KK Particle (LKP) stable \cite{stablekk}. KK-parity functions in a way which is analagous to R-parity in supersymmetric models, making it possible for the LKP to be a viable dark matter candidate \cite{kkdark}.

The identity of the LKP depends on the mass spectrum of the first KK level. The LKP is, most naturally, the first KK excitation of the $\B$. The KK spectrum to one-loop level is given in Ref.~\cite{CMS}. Using this spectrum, the relic density of the LKP can be calculated as a function of its mass. It has been found that the appropriate density is predicted when the mass 
is moderately heavy, between $600$ to $1200$ GeV \cite{kkdark}, somewhat heavier than the range favored in the case of supersymmetry. This range of the LKP mass depends on the details of the coannihilations of LKPs with heavier KK particles.

Another difference between dark matter particles in universal extra dimensions and supersymmetry is that unlike in the case of a neutralino LSP, the bosonic nature of the LKP 
means there is no chirality suppression of the annihilation signal
into fermions.  The annihilation rate of the LKP is therefore roughly 
proportional to the hypercharge$^4$ of the final state, leading 
to a large rate into leptons, including neutrinos. The annihilation and coannihilation cross sections are determined by Standard Model
couplings and the mass spectrum of the first KK level.

\section{Capture and Annihilation in the Sun}

In order to provide an observable flux of neutrinos, dark matter particles must be gathered in high concentrations. Deep gravitational wells such as the Sun, Earth or galactic center are examples of regions where such concentrations may be present. In the following calculation, we will focus on the Sun, as its prospects are the most promising.

The rate at which WIMPs are captured in the Sun depends on the nature of the interaction the WIMP undergoes with nucleons in the Sun. For spin-dependent interactions, the capture rate is given by \cite{capture} 
\begin{equation} 
C_{\mathrm{SD}}^{\odot} \simeq 3.35 \times 10^{20} \, \mathrm{s}^{-1} 
\left( \frac{\rho_{\mathrm{local}}}{0.3\, \mathrm{GeV}/\mathrm{cm}^3} \right) 
\left( \frac{270\, \mathrm{km/s}}{\bar{v}_{\mathrm{local}}} \right)^3  
\left( \frac{\sigma_{\mathrm{H, SD}}} {10^{-6}\, \mathrm{pb}} \right)
\left( \frac{100 \, \mathrm{GeV}}{m_{\rm{DM}}} \right)^2 
\label{c-eq}
\end{equation} 
where $\rho_{\mathrm{local}}$ is the local dark matter density, 
$\sigma_{\mathrm{H,SD}}$ is the spin-dependent, WIMP-on-proton (hydrogen)
elastic scattering cross section, $\bar{v}_{\mathrm{local}}$ 
is the local rms velocity of halo dark matter particles and 
$m_{\rm{DM}}$ is our dark matter candidate.  The analogous formula for the 
capture rate from spin-independent (scalar) scattering is \cite{capture}
\begin{equation}
C_{\mathrm{SI}}^{\odot} \simeq 1.24 \times 10^{20} \, \mathrm{s}^{-1} 
\left( \frac{\rho_{\mathrm{local}}}{0.3\, \mathrm{GeV}/\mathrm{cm}^3} \right) 
\left( \frac{270\, \mathrm{km/s}}{\bar{v}_{\mathrm{local}}} \right)^3 
\left( \frac{2.6 \, \sigma_{\mathrm{H, SI}}
+ 0.175 \, \sigma_{\mathrm{He, SI}}}{10^{-6} \, \mathrm{pb}} \right) 
\left( \frac{100 \, \mathrm{GeV}}{m_{\rm{DM}}} \right)^2 \; .
\end{equation}
Here, $\sigma_{\mathrm{H, SI}}$ is the spin-independent, WIMP-on-proton
elastic scattering cross section and $\sigma_{\mathrm{He, SI}}$ is the 
spin-independent, WIMP-on-helium elastic scattering cross section. 
Typically, $\sigma_{\mathrm{He, SI}} \simeq 16.0 \, \sigma_{\mathrm{H, SI}}$.
The factors of $2.6$ and $0.175$ include information on the solar 
abundances of elements, dynamical factors and form factor suppression.

Although these two rates appear to be comparable in magnitude, the spin-dependent and spin-independent cross sections can differ radically. For example, with Kaluza-Klein dark matter, the spin-dependent cross section is typically three to four orders of magnitude larger than the spin-independent cross section \cite{kkdark} and solar accretion by spin-dependent 
scattering dominates.

If the capture rates and annihilation cross sections are sufficiently
high, the Sun will reach equilibrium between these processes.  
For $N$ (number of) WIMPs in the Sun, the rate of change of this
number is given by
\begin{equation}
\dot{N} = C^{\odot} - A^{\odot} N^2 \; ,
\end{equation}
where $C^{\odot}$ is the capture rate and $A^{\odot}$ is the 
annihilation cross section times the relative WIMP velocity per volume.  
$C^{\odot}$ was given in (\ref{c-eq}), while $A^{\odot}$ is 
\begin{equation}
A^{\odot} = \frac{\langle \sigma v \rangle}{V_{\mathrm{eff}}} 
\end{equation}
where $V_{\mathrm{eff}}$ is the effective volume of the core
of the Sun determined roughly by matching the core temperature with 
the gravitational potential energy of a single WIMP at the core
radius.  This was found in Ref.~\cite{equ} to be
\begin{equation}
V_{\rm eff} = 5.7 \times 10^{27} \, \mathrm{cm}^3 
\left( \frac{100 \, \mathrm{GeV}}{m_{\rm{DM}}} \right)^{3/2} \; .
\end{equation}
The present WIMP annihilation rate is 
\begin{equation} 
\Gamma = \frac{1}{2} A^{\odot} N^2 = \frac{1}{2} \, C^{\odot} \, 
\tanh^2 \left( \sqrt{C^{\odot} A^{\odot}} \, t_{\odot} \right) \; 
\end{equation}
where $t_{\odot} \simeq 4.5$ billion years is the age of the solar system.
The annihilation rate is maximized when it reaches equilibrium with
the capture rate.  This occurs when 
\begin{equation}
\sqrt{C^{\odot} A^{\odot}} t_{\odot} \gg 1 \; .
\end{equation}
For the majority of particle physics models which are most often considered (most supersymmetry or Kaluza-Klein models, for example), the WIMP capture and annihilation rates reach, or nearly reach equilibrium in the Sun. This is often not the case for the Earth. This is true for two reasons. First, the Earth is less massive than the Sun and, therefore, provides fewer targets for WIMP scattering and a less deep gravitational well for capture. Second, the Earth accretes WIMPs only by scalar (spin-independent)
interactions. For these reasons, it is unlikely that the Earth will provide any observable neutrino signals from WIMP annihilations in any planned experiments.

If very high densities of dark matter are present in the galactic center, such as would be expected for very cuspy halo profiles \cite{cusp} or density spikes \cite{spike}, sizable neutrino fluxes could be produced. For most particle dark matter candidates, however, very large fluxes of gamma-rays would accompany such neutrinos and it would be unlikely that neutrinos would be observed in the absence of a gamma-ray signal. Neutrino experiments could, however, help confirm that a gamma-ray signal was the result of dark matter annihilations rather than a more traditional astrophysical source.

\section{Neutrinos and Their Detection}

The rate of neutrinos produced in WIMP annihilations is highly model dependent as the annihilation fractions to various products can vary a great deal from model to model. We will attempt to be as general in our discussion as possible while still considering some specific cases.

In supersymmetry, there are no tree level diagrams for direct neutralino annihilation to neutrinos. Many indirect channels exist, however. These include neutrinos from heavy quarks, gauge bosons, tau leptons and higgs bosons. These processes result in a broad spectrum of neutrinos, but with typical energies of 1/2 to 1/3 of the neutralino mass. For experimental energy thresholds of 10-100 GeV, lighter WIMPs can be very difficult or impossible to detect for this reason.

For neutralinos lighter than the $W^{\pm}$ mass ($80.4\,\,$GeV), annihilation to $b \bar{b}$ typically dominates, with a small admixture of $\tau^+ \tau^-$ as well. In these cases, neutrinos with less than about 30 GeV energy are produced and detection is difficult. For heavier neutralinos, annihilation into gauge bosons, top quarks and higgs bosons are important in addition to $b \bar{b}$ and $\tau^+ \tau^-$. In particular, gauge bosons can undergo two body decay ($Z \rightarrow \nu \nu\,\,$ or $W^{\pm} \rightarrow l^{\pm} \nu$) producing neutrinos with an energy of about half of the WIMP mass. Neutralinos with substantial higgsino components often annihilate mostly into gauge bosons. 

For Kaluza-Klein dark matter, the picture is somewhat different. The LKP annihilates directly to a pair of neutrinos about 3-4\% of the time \cite{kkdark,kribs}. Although this fraction is small, the neutrinos are of higher energy and are, therefore, easier to detect. The more frequent annihilation channels for Kaluza-Klein dark matter are charged leptons (60-70\%) and up-type quarks (20-30\%). Of these, the $\tau^+ \tau^-$ mode contributes the most to the neutrino flux. Unlike in supersymmetry, a large fraction of LKPs annihilate into long lived particles, such as up quarks, electrons and muons, which lose their energy in the Sun long before decaying. Bottom and charm quarks lose some energy before decaying, but not as dramatically.

Neutrinos which are produced lose energy as they travel through the Sun \cite{edsjo,JK,crotty}. The probability of a neutrino escaping the sun without interaction is given by \cite{crotty}
\begin{equation} 
P = e^{-E_{\nu}/E_k}
\end{equation}
where $E_k$ is $\simeq130$ GeV for $\nu_{\mu}$, $\simeq160$ GeV for
$\nu_{\tau}$, $\simeq200$ GeV for $\bar{\nu_{\mu}}$ and $\simeq230$ GeV for
$\bar{\nu_{\tau}}$. Thus we see that neutrinos above a couple of  hundred GeV are especially depleted, although those which escape are also more easily detected. For a useful parameterization of solar effects, see Ref.~\cite{edsjo}. Note that neutrino oscillations can also play an important role in calculating the flux of muon neutrinos in a detector \cite{crotty}.

A small fraction of the muon neutrinos which reach the detector are converted 
to muons through charged current interactions \cite{crosssection}. These muons then propagate through the Cerenkov medium of the detector, where they are detected by photo-multiplier tubes distributed through the effective volume. For a review of high-energy neutrino astronomy, see Ref.~\cite{review}.

As a muon propagates, it loses energy at the rate
\begin{equation}
\frac{dE}{dX} = -\alpha - \beta E
\end{equation}
where $\alpha= 2.0 \ \rm{MeV} \, \rm{cm}^2/\rm{g}$ and $\beta= 4.2 \times 10^{-6} \,\rm{cm}^2/\rm{g}$. The distance a muon travels before dropping below the threshold energy, called the muon range, is then given by Ref.~\cite{range}
\begin{equation} 
R_{\mu} \simeq \frac{1}{\rho \beta} \mathrm{ln}\bigg[\frac{\alpha + \beta E_{\mu}}{\alpha+\beta E_{\mathrm{thr}}} \bigg]
\end{equation}
where $\rho$ is the density of the detector medium, $\alpha \simeq 2.0$
MeV cm$^2$/g and $\beta \simeq 4.2 \times 10^{-6}$ cm$^2$/g. $E_{\mu}^{\rm{thr}}$ is the energy threshold of the detector, typically 10-100 GeV for deep ice or water detectors. The effective volume in which a muon producing interaction can occur and be observed is simply the muon range times the effective area of the detector.

Currently, the AMANDA experiment is taking data at the South Pole \cite{amanda}. With an effective area of 50,000 square meters and a muon energy threshold of about 30 GeV, AMANDA is currently the largest volume high-energy neutrino telescope. Due to the ``soda can'' geometry of AMANDA B-10, it was not very sensitive in the direction of the Sun (the horizon), and could only place useful limits on neutrinos from the center of the Earth. As we said before, the rate of dark matter annihilations in the Earth is much smaller than in the Sun, making detections unlikely. The current version of this experiment, AMANDA-II, does not have this problem, however, and is sensitive to neutrinos from both the Sun and Earth.

ANTARES \cite{antares}, now under construction in the Mediterranean, will have a similar effective area as AMANDA-II, but with a lower energy threshold (10 GeV). ANTARES may be an important dark matter experiment, especially if WIMPs are rather light. Also, unlike neutrino telescopes at the South Pole, ANTARES will also be sensitive to neutrinos from the galactic center. 

IceCube \cite{icecube}, under construction at the South Pole, will have a full square kilometer of effective area, but with a somewhat higher energy threshold (50-100 GeV). 

The background for this class of experiments consists of atmospheric
neutrinos \cite{atmback} and neutrinos generated in cosmic ray
interactions in the Sun's corona \cite{sunback}.  In the direction of the
Sun (up to the angular resolution of a neutrino telescope), tens of events
above 100 GeV and on the order of 1 event per year above 1 TeV per square
kilometer are expected from the atmospheric neutrino flux.  Fortunately,
for a very large volume detector with sufficient statistics, this
background is expected to be significantly reduced, and possibly 
eliminated.  Furthermore, this rate could be estimated based on the 
rate from atmospheric neutrinos, a level of about a few events per year. 
The final background is then further reduced by selecting on judiciously
chosen angular and/or energy bins.  Neutrinos generated by cosmic ray 
interactions in the Sun's corona, however, cannot be reduced in this way.
This irreducible background is predicted to be less than a few events per
year per square kilometer above 100 GeV.

\section{Prospects}

The current limits on dark matter annihilation from AMANDA and other similar experiments are not very strong. Experiments with lower energy thresholds (ANTARES), larger effective areas (IceCube) and which can observe in the direction of the Sun (ANTARES and IceCube) will greatly enhance this sensitivity in coming years.

\begin{figure}[t]
\centering\leavevmode
\includegraphics[width=3.5in]{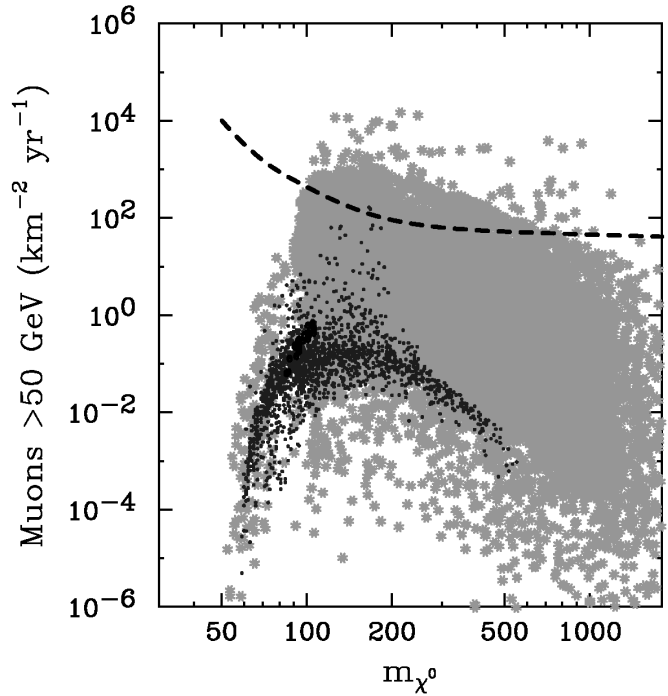}
\caption{The number of events from neutralino annihilation in the Sun per year in a detector with effective area equal to one square kilometer and a 50 GeV muon threshold \cite{wang}. The lightly shaded region represents the general Minimal Supersymmetric Standard Model (MSSM), the darker region corresponds to mSUGRA models, a subset of the MSSM. For each point shown, the relic density is below the maximum value allowed by the WMAP data ($\Omega_{\chi} h^2 \le 0.129$). The sensitivity projected for IceCube is shown as a dashed line \cite{edsjolimits}.}
\label{figsusy1}
\end{figure}

The sensitivity of a square kilometer neutrino detector with a moderate muon energy threshold (50 GeV) to supersymmetric dark matter is shown in Fig.~\ref{figsusy1}. From this figure, it is clear that high-energy neutrinos will be an observable signature in only a small fraction of possible supersymmetry models, although such experiments are still certainly an important probe.

\begin{figure}[t]
\centering\leavevmode
\mbox{
\includegraphics[width=3.2in]{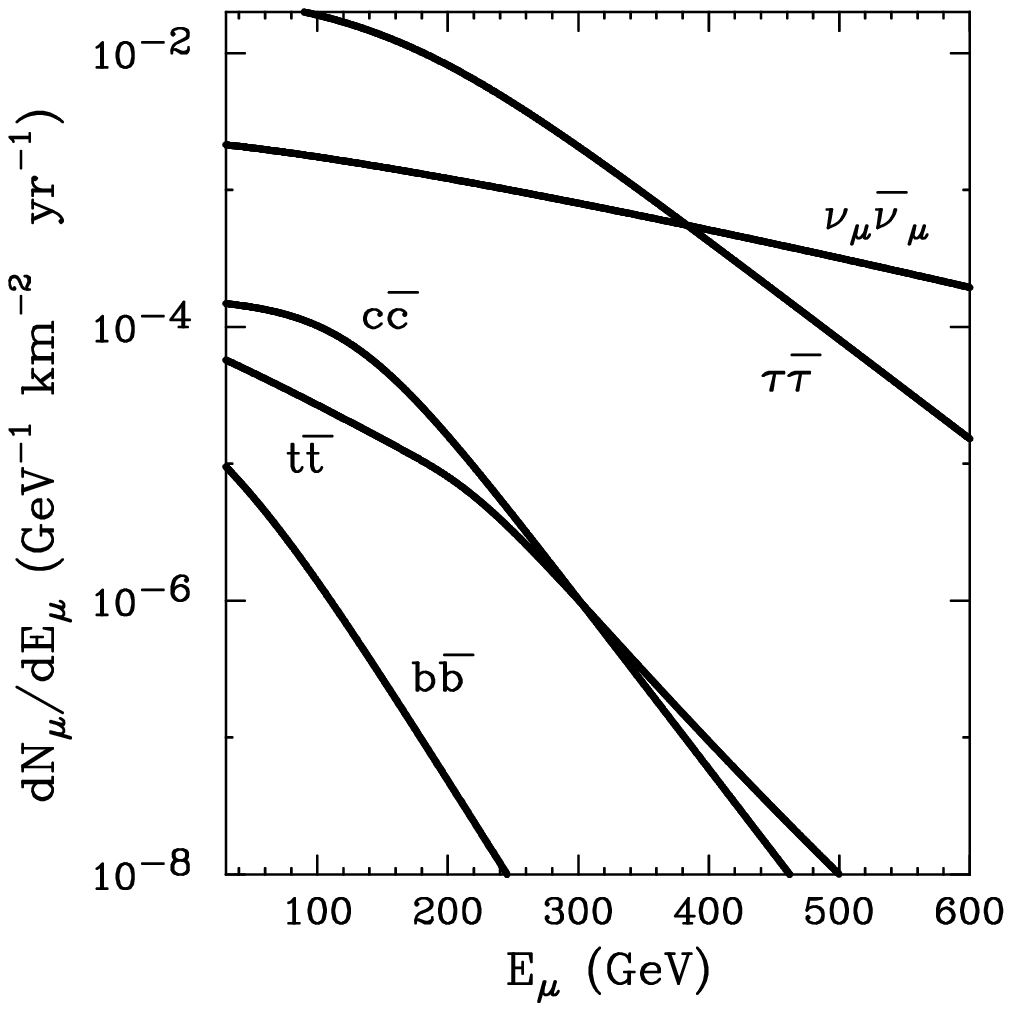}
\hfill
\includegraphics[width=3.2in]{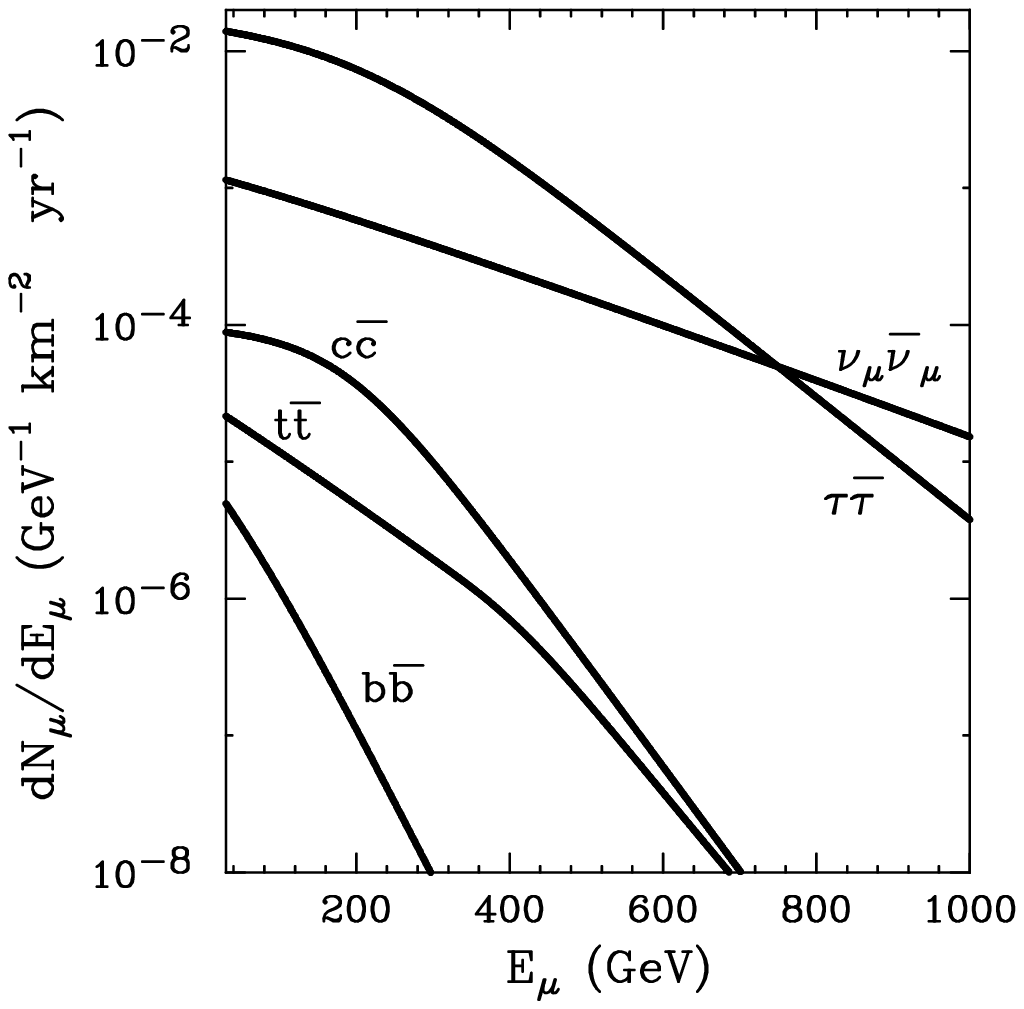}}
\caption{The spectrum of muons at the Earth generated in charged-current 
interactions of muon neutrinos generated in the annihilation of 
$600$ GeV (left side) and $1000$ GeV (right side)
dark matter particles in the Sun \cite{kribs}.  The elastic scattering cross section used for capture in the Sun was fixed at $10^{-6}$ pb for both graphs.  
The rates are proportional to that cross section.}
\label{figkk1}
\end{figure}

For Kaluza-Klein dark matter, the prospects for detection via high-energy neutrinos are substantially better. This is largely because of which annihilation modes dominate. The spectrum of muons in a detector on Earth due to LKP annihilations in the Sun is shown in Fig.~\ref{figkk1} for various annihilation channels and for two choices of LKP mass. Unlike in the case of supersymmetry, annihilation to neutrinos and taus dominate the neutrino spectrum. In supersymmetry, b quarks and gauge bosons dominate, producing fewer observable neutrinos.

\begin{figure}[t]
\centering\leavevmode
\includegraphics[width=3.5in]{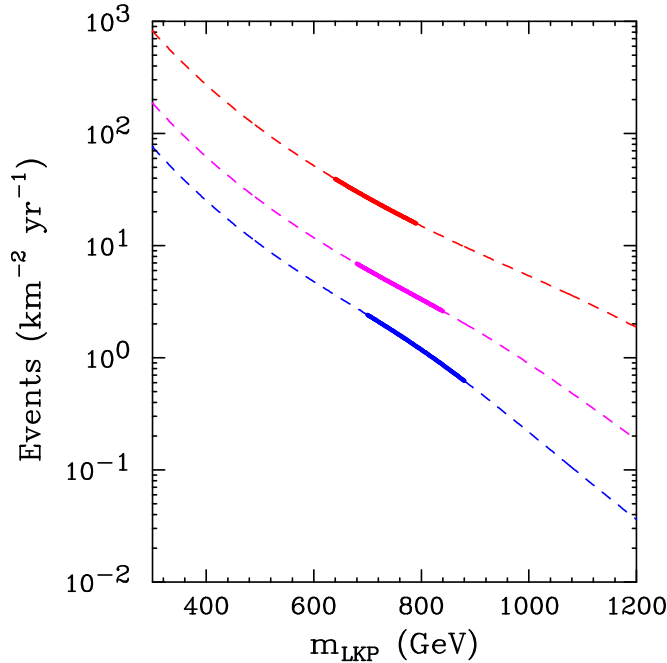}
\caption{The number of events per year from Kaluza-Klein dark matter annihilation in the Sun in a detector with an effective 
area equal to one square kilometer and a muon energy threshold of 50 GeV \cite{kribs}.  Contours are shown, from bottom to top, for 
$r_{\qR}=0.1$, $0.2$, and $0.3$, where $r_{\qR}$ is the mass splitting of the LKP and the next-to-LKP over the LKP mass. The 
expected size of the one-loop radiative corrections predict $0.1 \lsim r_{\qR} \lsim 0.2$, therefore, the $r_{\qR}=0.3$ contour
is shown merely for comparison.
The relic density of the LKP's lies within the range
$\Omega_{\B} h^2 = 0.16 \pm 0.04$ for the solid sections 
of each line.  The relic density is smaller (larger) for 
smaller (larger) LKP masses.}
\label{figkk2}
\end{figure}

In Fig.~\ref{figkk2}, the event rates in a square kilometer detector are shown (using a threshold of 50 GeV). Each of the three lines corresponds to a 
different value of the next-to-lightest KK particle's mass. The 
expected size of the one-loop radiative corrections predicts $0.1 \lsim r_{\qR} \lsim 0.2$, where $r_{\qR}$ is the mass splitting of the LKP and the next-to-LKP over the LKP mass.  For this range,
a kilometer scale neutrino telescope would be sensitive to a 
LKP with mass up to about 1 TeV.  The relic density
of the LKP varies from low to high values from 
left to right in the graph.  The range of mass of the LKP that
gives the appropriate relic density was estimated from Ref.~\cite{kkdark} and is shown in the
figure by the solid sections of the lines. Combining the expected
size of the one-loop radiative corrections with a relic density
appropriate for dark matter, we see that IceCube should see 
between a few to tens of events per year.  

For detectors with smaller effective areas one simply has to scale the 
curves down by a factor $A/(1 \; \mathrm{km}^2)$ to obtain the event rate.  
In particular, for the first generation neutrino telescopes including AMANDA, 
ANTARES, and NESTOR, with effective areas of order $0.1$ km$^2$,
the event rate could be as high as a few events per year for a
LKP mass at the lower end of the solid line region.

\section*{Acknowledgments}

DH is supported by the Leverhulme trust.

\end{document}